\documentclass[doublecol]{epl2} 

\title{Recovery of state-specific potential of molecular motor from single-molecule trajectory}
\shorttitle{Recovery of molecular-motor's potential} %Insert here a short version of the title if it exceeds 70 characters

\author{Shoichi Toyabe\inst{1,2} \and Hiroshi Ueno\inst{1} \and Eiro Muneyuki\inst{1}\thanks{E-mail: \email{emuneyuk@phys.chuo-u.ac.jp}}}
\shortauthor{S. Toyabe \etal}

\institute{                    
  \inst{1} Faculty of Science and Engineering, Chuo University, Tokyo 112-8551, Japan\\
  \inst{2} Faculty of Physics, Ludwig-Maximilians-Universit{\"a}t M{\"u}nchen 80339, Germany.
}


\newcommand{\PP}{P$_\mathrm{i}$\,}

\newcommand{\argmax}{\mathop{\rm argmax} \limits}

\abstract{
We have developed a novel method to evaluate the potential profile of a molecular motor at each chemical state from only the probe's trajectory and applied it to a rotary molecular motor F$_1$-ATPase. 
By using this method, we could also obtain the information regarding the mechanochemical coupling and energetics.
We demonstrate that the position-dependent transition of the chemical states is the key feature for the highly efficient free-energy transduction by F$_1$-ATPase.
}

\begin{document}

\maketitle

A molecular motor is a chemical engine that converts chemical free energy into mechanical motion; to understand the design principles of the molecular motor, it is essential to reveal its energetics.
During the conversion, chemical potential is converted into mechanical potential that then powers the motor's mechanical motion.
Thus far, for an ATP-driven rotary molecular motor, F$_1$-ATPase, or F$_1$-motor \cite{Boyer1993ah, Abrahams1994, Noji1997}, the outputs of the energy transduction such as the heat dissipation through rotations\cite{Yasuda1998, Toyabe2010a, Hayashi2010}, and its maximum work per step against external torque\cite{Toyabe2011} have been measured.
These studies have revealed that the F$_1$-motor can transduce the free energy change of ATP hydrolysis to mechanical motion at nearly 100\% efficiency.
However, the mechanism to achieve such a high efficiency remains unclear.
To reveal this, information about the mechanical potential that drives the $\gamma$-shaft's rotation is crucial.

\begin{figure}[tbp] 
\centerline{\includegraphics{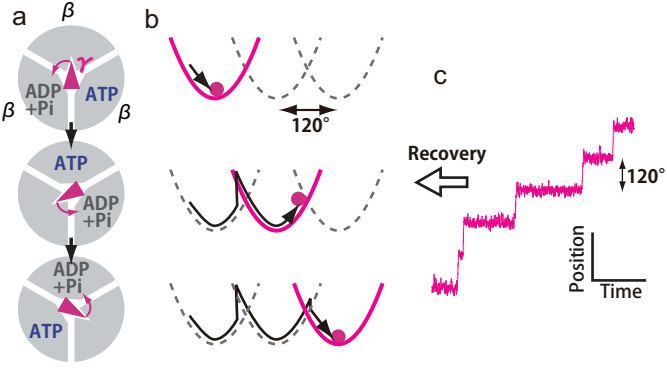}}
\caption{\label{fig:Scheme}%
Recovery of potentials from single-molecule trajectories.
{\bf a}, Simplified reaction scheme of F$_1$-motor.
$\alpha$ subunit is omitted for simplicity. 
The binding of ATP to the empty $\beta$ subunit triggers 120$^\circ$ rotation.
{\bf b}, Interaction potentials between the $\gamma$ shaft and $\beta$ subunits.
{\bf c}, Single-molecule trajectory. We recover the potential profiles ({\bf b}) from single-molecule trajectories ({\bf c}).
}
\end{figure}

In this letter, we describe our new method for recovering the mechanical potential profile of each chemical state only from the single-molecule trajectory (Fig. \ref{fig:Scheme}c) and its application to the rotational trajectory of the F$_1$-motor.
By using this method, we could also obtain information regarding the mechanochemical coupling and energetics.
The method is simple and applicable to not only other molecular motors but also systems in a broad range.

\begin{figure}[htb] % ŠT"O}
\centerline{\includegraphics{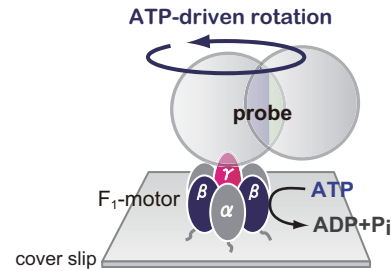}}
\caption{\label{fig:F1-motor}%
F$_1$-motor ($\alpha_3\beta_3\gamma$ sub complex).
The $\gamma$ shaft rotates unidirectionally when ATP is hydrolyzed to ADP and P$_\mathrm{i}$.
By attaching the $\alpha_3\beta_3$ ring to a glass cover slip and a dimeric probe particle to the $\gamma$ shaft, the rotation can be observed under an optical microscope.
}
\end{figure}

F$_1$-motor is a rotary molecular motor, whose central $\gamma$ shaft rotates uni-directionally against a stator ring ($\alpha_3\beta_3$) while hydrolyzing ATP to ADP and phosphate (P$_\mathrm{i}$) (Fig. \ref{fig:F1-motor}a). 
At low ATP concentrations, F$_1$-motor exhibits a stepwise rotation with pauses corresponding to ATP waiting states.
During this state, each $\beta$ subunit has one of the following three different chemical states: with ATP, with ADP (and P$_\mathrm{i}$), and empty \cite{Adachi2007, Watanabe2010} (Fig. \ref{fig:Scheme}a).
The chemical states of the three $\beta$ subunits rotate cooperatively due to the ATP binding, ATP hydrolysis, and the release of ADP and a phosphate.
Since the conformation of the $\beta$ subunit varies with its chemical state\cite{Abrahams1994, Masaike2008}, the interaction potential between the $\beta$ subunits and $\gamma$ shaft rotates and drives the $\gamma$-shaft's 120$^\circ$ rotation. 
This picture leads to a simple model that the stepwise rotation is described as a Brownian motion on a potential corresponding to the ATP waiting state, wherein the potential's position shifts 120$^\circ$ upon the ATP binding (Fig. \ref{fig:Scheme}b).
Such discrete model is known to well describe the F$_1$-motor's stepwise rotation\cite{Gaspard2007}.

\section{Analytical methods}

Let $X$ and $S$ denote collectively the observed trajectory of the motor's probe $(x_1,\dots, x_\mathrm{L})$ and the chemical-state sequence of the motor $(s_1,\dots, s_\mathrm{L-1};s_\mathrm{k}\in \{1, \ldots, N\})$, respectively, where $L$ is the number of the video frames and $N$ is the number of states.
In the case of F$_1$-motor at low ATP concentrations, $N=3$.
The essence of our method is to find the most probable chemical-state sequence $S^*$ that maximizes a well-defined score function $\sigma(S, X)$ for the given $X$:
\begin{equation}
S^*=\argmax_S\sigma(S, X).
\end{equation}
We define the score function as the joint path probability of $X$ and $S$: $\sigma(X, S)=\sigma_\mathrm{S}(X)\sigma(S)$.
Here,
\begin{equation}
\sigma(S)\equiv P(s_1)\prod_{k=1}^{L-2}W(s_\mathrm{k+1}|s_\mathrm{k})
\end{equation}
is the Markovian path probability that $S$ realizes, $P(s)$ is the probability that a state $s$ realizes, and $W(s'|s)$ is the transition probability from a state $s$ to another state $s'$.
On the other hand,
\begin{equation}
\sigma_\mathrm{S}(X)\equiv p_\mathrm{s_1}(x_1)\prod_{k=1}^{L-1}w_\mathrm{s_k}(x_\mathrm{k+1}|x_\mathrm{k})
\end{equation}
is the Markovian path probability that $X$ realizes under a given $S$; here,  $p_\mathrm{s}(x)$ is the probability of observing $x$ at a state $s$, and $w_\mathrm{s}(x'|x)$ is the transition probability from $x$ to $x'$ in successive two frames at $s$.
Let $D$ and $\tau$ denote the diffusion coefficient and the sampling period, respectively.
When the mean square displacement between successive frames $D\tau$ is sufficiently small, $w_{s}(x'|x)$ becomes approximately
\begin{equation}
\frac 1{\sqrt{4\pi D\tau}}\exp\left[-\left(x'-x+\frac{D\tau}{k_\mathrm{B}T}\partial_x U_\mathrm{s}(x)\right)^2/4D\tau\right]
\end{equation}
to the first order of $D\tau$ \cite{Risken},
where $k_\mathrm{B}T$ is the thermal energy and $U_\mathrm{s}(x)$ is the mechanical potential at a state $s$.

\begin{table}[t]
\caption{List of symbols.}
\begin{tabular}{l|l}
\hline $L$ & Number of video frames \\ 
\hline $N$ & Number of chemical states \\ 
\hline $X$ & Trajectory of the motor's position, \\
& $(x_1,\dots, x_\mathrm{L})$ \\ 
\hline $S$ & Chemical-state sequence, $(s_1,\dots, s_\mathrm{L})$ \\ 
\hline $S^*$ & $\argmax_S \sigma(S, X)$.     \\ 
%&  Chemical-state sequence that maximizes  \\ 
%& the score function. \\
\hline $\sigma(X, S)$ & Score function to be maximized\\
\hline $\sigma_\mathrm{S}(X)$ & Path probability that $X$ realizes under $S$ \\ 
\hline $\sigma(S)$ & Path probability that $S$ realizes \\ 
\hline $p_\mathrm{s}(x)$ & Probability that $x$ realizes at $s$ \\ 
\hline $P(s)$ & Probability that $s$ realizes \\ 
\hline $w_\mathrm{s}(x'|x)$ & Transition probability from $x$ to $x'$ at $s$ \\ 
\hline $W(s'|s)$ & Transition probability from $s$ to $s'$ \\ 
\hline $D$ & Diffusion coefficient \\ 
\hline $U_\mathrm{s}(x)$ & Potential profile at $s$ \\ 
\hline $\tau$ & Sampling period \\ 
\hline 
\end{tabular} 
\end{table}

In the below simulations and experiments, we calculated $D$ as the mean power spectrum density of the rotational rate in the frequency range from 0.8 kHz to 1.2 kHz since the fluctuation-dissipation relation is expected to hold at the high frequency region\cite{Toyabe2010a}.
For $p_\mathrm{s}(x)$, we used Gaussian functions fitted to the peaks of the probe's angular distribution.
For evaluating $\sigma(X, S)$, we need to know $U_\mathrm{s}(x)$, $P(s)$, and $W(s'|s)$ additionally, which are not known {\it a priori}.
We iteratively evaluate these unknown parameters as well as $S^*$ as follows.

The algorithm consists of three steps. 
On the basis of rough estimations of the parameters (Step 1), we find $S^*=\argmax_S\sigma(X, S)$ by using Viterbi algorithm \cite{NumericalRecipes3} (Step 2).
Then, we update $U_\mathrm{s}(x)$, $P(s)$, and $W(s'|s)$ on the basis of $X$ and $S^*$ (Step 3).
We iterate Step 2 and Step 3 alternatively typically 20 times. %, we evaluate $S^*$, $U_\mathrm{s}(x)$, $P(s)$, and $W(s'|s)$.
$\sigma(X, S^*)$ does not always increase monotonically over iterations.
We adopt $S^*$, $U_\mathrm{s}(x)$, $P(s)$, and $W(s'|s)$ that maximize $\sigma(X, S^*)$ as the final estimation.
The detailed procedures are as follows.

{\it Step 1: Initialization.}
As the initial estimation, we use $U_\mathrm{s}(x)=-\ln p_\mathrm{s}(x)$, $P(s)=1/N$, and $W(s'|s)=M/L$, where $M$ is the total number of steps estimated from the trajectory.

{\it Step 2: State identification.}
We find $S^*$ that rigorously maximizes $\sigma(X, S)$.
This seems hopeless since the number of possible $S$ is $N^L$ ($N=3$ and $L\sim 2,500,000$ in the experiment below).
However, the Viterbi algorithm finds $S^*$  at a computational cost of $O(N^2L)$, which depends on $L$ linearly \cite{NumericalRecipes3}.

{\it Step 3: Potential estimation and parameter update.}
We divide the spatial coordinate to bins and calculate the number of transitions between the bins for every consecutive two frames.
Let $n_\mathrm{i\to j}$ be the number of transitions from the $i$-th bin to the $j$-th bin at a state.
The equilibrium distribution at this state, $\{\pi_\mathrm{i}\}$, satisfies the balance condition, $\pi_\mathrm{j}=\sum_i\omega_\mathrm{i\to j}\pi_\mathrm{i}$, where $\omega_\mathrm{i\to j}\equiv n_\mathrm{i\to j}/\sum_kn_\mathrm{i\to k}$.
By solving this equation, we obtain the potential profile in this state as $U_\mathrm{i}=-k_\mathrm{B}T\ln \pi_\mathrm{i}$ with a trivial additive constant.
We fit $U_\mathrm{i}$ by using a sixth-order polynomial function for estimating $U_\mathrm{s}(x)$.
We update $P(s)$ and $W(s'|s)$ to $P(s)=m_\mathrm{s}/(L-1)$ and $W(s'|s)=m_\mathrm{s\to s'}/m_\mathrm{s}$, where $m_\mathrm{s}$ and $m_\mathrm{s\to s'}$ are the numbers of $s$ and steps from $s$ to $s'$ included in $S^*$, respectively.

\section{Numerical simulation}

\begin{figure}[ht]
\centerline{\includegraphics{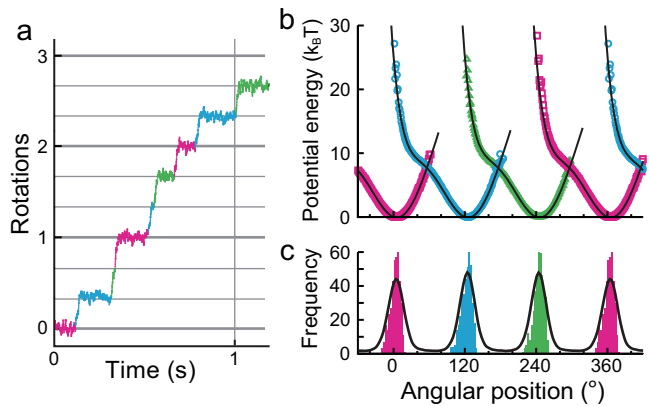}}
\caption{\label{fig:Numerical computation}%
Numerical simulation of the Langevin equation, $\gamma \dot x=-\partial_xU(x(t)-x_\mathrm{0}(t))+\xi(t)$;
$\langle\xi(t)\rangle=0$; $\langle\xi(0)\xi(t)\rangle=2\gamma k_\mathrm{B}T\delta(t)$.
Here, $\gamma= 0.3\,\mathrm{pN\,nm\,s/rad^2}$ is the rotational frictional coefficient ($D=13.8\, \mathrm{rad^2/s}$), $T= 300\, \mathrm{K}$, the frame rate is 9,000 Hz, and the frame length is $2\times 10^6$.
$U(x)=0.5x^6-0.5x^5-3x^4+3x^3+10x^2-2x$, where $x$ is expressed in radians.
The position of the potential $x_0$ is shifted with $120^\circ$ spacings unidirectionally randomly in a Poissonian process at a rate of 6 Hz.
These parameters are chosen to be consistent with the experiment.
Transitions within 20 ms after the previous transition are prohibited since such successive transitions are not expected in experiments.
{\bf a}, Numerically generated trajectory coloured according to the estimated states.
{\bf b}, Recovered potential profiles (symbols) and the theoretical profile $U(x)$ (solid lines).
{\bf c}, Angular distribution at transitions (histogram) and the overall angular distribution (solid line).
The overall angular distribution is scaled by eye to be compared with the transition angular distributions.
The potential profile in {\bf b} and angular distribution at transitions from this state in {\bf c} are depicted using the same colour.
}
\end{figure}

The validity of this method was tested by employing a simulated trajectory of a Brownian particle.
The particle was assumed to be constrained in a non-harmonic potential and obey a Langevin equation with parameters consistent with the below experiment (Fig. \ref{fig:Numerical computation}a).
The potential's position was shifted with 120$^\circ$ spacings randomly in a Poissonian process.
Figure \ref{fig:Numerical computation}b demonstrates that our method recovered the potential profile of each state accurately  only from the stepwise trajectory.
By using this method, we can also evaluate the angular positions where the transitions occur ($\theta_\mathrm{tr}$); if $s^*_\mathrm{k}\ne s^*_\mathrm{k+1}$ for a certain $k$, $x_\mathrm{k+1}$ is the transition position for this transition.
The distribution of $\theta_\mathrm{tr}$ was similar to the overall angular distribution (Fig. \ref{fig:Numerical computation}c).
This correctly reflects the fact that we shifted the states randomly.
The estimation error of $\theta_\mathrm{tr}$ was $0.3 \pm 9.8^\circ$ (mean $\pm$ SD).
$D$ was estimated to be 13.9 $\mathrm{rad^2/s}$ in good agreement with the simulation parameter (13.8 $\mathrm{rad^2/s}$).

\section{Experiment}

\begin{figure}[ht]
\centerline{\includegraphics{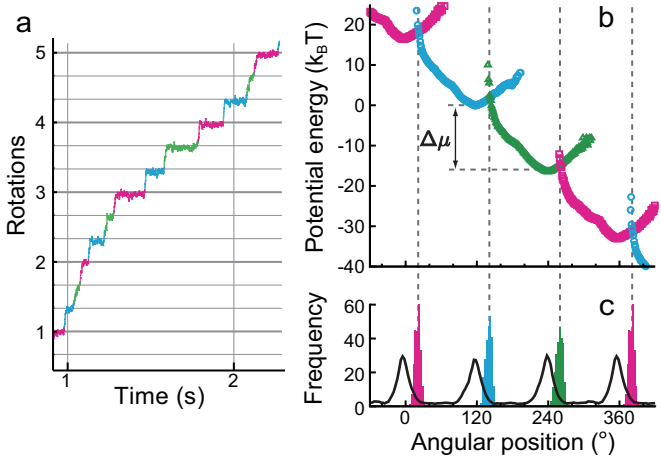}}
\caption{\label{fig:Result}
Experiment.
Recovered free-energy potentials and the angular distribution at transitions from the trajectory of an F$_1$-motor molecule.
{\bf a}, Rotational trajectory of the F$_1$-motor's probe particle observed at 9,000 Hz with an exposure time of 40 $\mu$s.
The trajectory comprises approximately 650,000 frames.
{\bf b}, Recovered potentials corresponding to the ATP-waiting states. The vertical lines denote the intersection points of the potentials.
The potentials are plotted with vertical shifts with the amount of the chemical free energy change of ATP hydrolysis ($\Delta\mu=16.5\,k_\mathrm{B}T$).
{\bf c}, Angular distribution at the transitions (histogram) and the overall angular distributions (solid line).
See the caption of Fig. \ref{fig:Numerical computation} for details.
}
\end{figure}

We applied the method to the rotational trajectory of the F$_1$-motor at a low ATP concentration (0.5 $\mu$M ATP, 2.5 $\mu$M ADP, and 1 mM P$_\mathrm{i}$) wherein the ATP binding was the rate-limiting step.
The $\gamma$-shaft's rotation was probed by a 300 nm dimeric particle and observed at 9,000 Hz with an exposure time of 40 $\mu$s by using a high-speed camera  (Fig. \ref{fig:Result}a).
%The probe attached to the $\gamma$ shaft was observed at 9,000 Hz with an exposure time of 40 $\mu$s by using a high-speed camera.
Such short exposure times provide information regarding instantaneous angular positions. 
Since backward steps are rarely observed in the absence of external loads\cite{Toyabe2011}, we restricted the analysis to only the forward steps (ATP-hydrolytic direction).
Note that our method can be applied to systems with bidirectional steps as well.

\begin{figure}[htb]
\centerline{\includegraphics{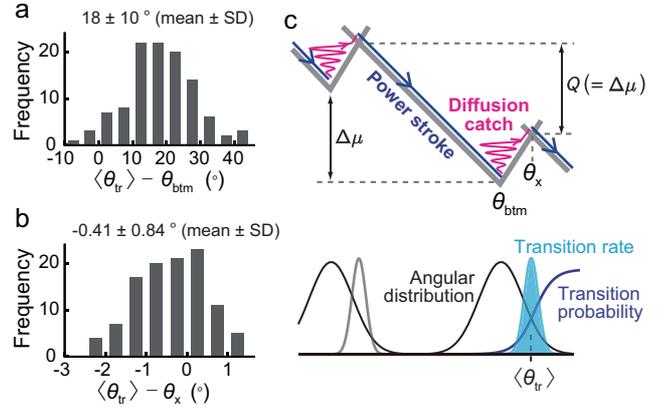}}
\caption{\label{fig:Discussion}
The distributions of $\langle\theta_\mathrm{tr}\rangle-\theta_\mathrm{btm}$ ({\bf a}) and $\langle\theta_\mathrm{tr}\rangle-\theta_\mathrm{x}$ ({\bf b}). 
36 molecules (108 states).
{\bf c}, The torque generation by the F$_1$-motor.
The heat through the rotations, $Q$, is calculated as the potential change between successive transitions. 
When the transition is limited around $\theta_\mathrm{x}$, $\Delta\mu\simeq Q$.
}
\end{figure}

Figure \ref{fig:Result}b shows the potential profiles corresponding to the three ATP-waiting states recovered from the single-molecule trajectories.
The potential had a simple structure with a single minimum.
The potentials are plotted with vertical shifts with the amount of the chemical free energy change of ATP hydrolysis ($\Delta\mu=16.5 k_\mathrm{B}T$, see {\it Materials and methods}).
We found that the transitions do not occur at random positions but only at a limited angular region ahead of the potential minimum in the rotational direction (Fig. \ref{fig:Result}c).
It is noteworthy that the potentials intersect around this region.
See Figs. \ref{fig:Discussion}a and b for the distributions of the mean transition angle ($\langle\theta_\mathrm{tr}\rangle$) relative to the potential minimum ($\theta_\mathrm{btm}$) and the intersection point of the potentials ($\theta_\mathrm{x}$), respectively.

It is expected that ATP frequently attacks the $\beta$ subunit irrespective of the $\gamma$-shaft's angular position.
However, our results suggest that the $\gamma$-shaft has to be ahead of the potential minimum for the ATP binding to proceed (Fig. \ref{fig:Discussion}c); otherwise, the attack fails.
Such a position-dependent substrate binding supports the ``diffusion catch'' mechanism for the torque generation \cite{Adachi2010, Toyabe2010b}, in which the thermally activated conformation suitable for the substrate binding is stabilized by the binding.
On the other hand, after the transition, the $\gamma$-shaft rotates along the potential of the new state into its minimum driven by the bending motion of the $\beta$-subunit (power stroke).
Thus, the mechanochemical picture of the torque generation was elucidated in terms of the basic concepts only from the trajectory.
Although we have assumed $W$ to be independent of the position for simplicity, the estimated transition probability depended on the position.
It is straightforward to use a position-dependent $W$; this would increase the reliability of the estimation.

Furthermore, by using this method, we can evaluate thermodynamic quantities such as the heat dissipation of the probe only from the trajectory. 
The heat dissipated to the environment through the rotational degree of freedom of the probe per step, $Q$, is equal to the potential change between successive transitions \cite{Sekimoto2010}.
Our result, $\langle\theta_\mathrm{tr}\rangle\simeq \theta_\mathrm{x}$, implies 
\begin{equation}\label{eq:Q}
Q\simeq \Delta\mu
\end{equation}
(Fig. \ref{fig:Discussion}c).
In fact, the estimated value of $Q$ was $17.8\pm 1.6\,k_\mathrm{B}T$ (mean $\pm$ SD) and was similar to $\Delta\mu$ ($16.5\,k_\mathrm{B}T$).
The slight difference between $Q$ and $\Delta\mu$ might be due to the estimation error of $\langle\theta_\mathrm{tr}\rangle$.
Because of the steep slope in the left-hand side of the potential (Fig. \ref{fig:Result}b), a small estimation error of $\langle\theta_\mathrm{tr}\rangle$ results in a large estimation error of $Q$.
It is possible that $\langle\theta_\mathrm{tr}\rangle$ estimated from the probe's motion is slightly different from that of the $\gamma$-shaft and biased toward the potential minimum due to the time delay of the probe's motion.
If this is the case, $\langle\theta_\mathrm{tr}\rangle$ is underestimated and therefore $Q$ is overestimated.
On the other hand, because of this steep potential, when $\Delta\mu$ is varied by controlling the concentrations of ATP, ADP, and P$_\mathrm{i}$, the changes in $\theta_\mathrm{x}$ are small.
Therefore, the relation (\ref{eq:Q}) is expected to hold over a broad range of $\Delta\mu$ provided reasonably that $\langle\theta_\mathrm{tr}\rangle$ does not depend on $\Delta\mu$ largely.
This suggests that F$_1$-motor transduces most $\Delta\mu$ to rotational motions at a nearly 100\% efficiency\cite{Toyabe2010a}.
This is consistent with a previous study\cite{Toyabe2010a}, wherein we had measured $Q$ from the violation of the fluctuation-dissipation relation using Harada-Sasa equality\cite{Harada2005}.
It required the response measurement against an external load, which is accessible only in limited systems.
The present method extracted a thermodynamic relation (\ref{eq:Q}) only from the trajectory and furthermore demonstrated that the position-dependent substrate binding, $\langle\theta_\mathrm{tr}\rangle\simeq \theta_\mathrm{x}$, is the mechanism behind this relation.

\section{Discussions}

The 120$^\circ$ step triggered by the ATP binding is known to comprise 80$^\circ$ and 40$^\circ$ sub steps with a short pause after the 80$^\circ$ sub step, in which ATP hydrolysis and phosphate release occur \cite{Yasuda2001, Shimabukuro2003, Masaike2008}.
The duration of this pause is too short to be resolved by the relatively large probes used here.
Therefore, the recovered potential is considered to be the effective one that the substate's potential is superimposed\cite{Prost1994, Wang2003}.
In other words, without knowing the details of the system, we can recover the coarse-grained potentials to the extent of the model description.
It is the next challenge to use an essentially friction-free probe such as a colloidal gold particle with tens of nanometre diameter \cite{Yasuda2001, Ueno2010} for better estimation as well as the recovery of the potential profiles during the catalytic dwells.

Recent developments in experimental techniques have enabled us to obtain single-molecule trajectories with high resolutions in both time and space.
However, the trajectory analyses have thus far been limited to quantities such as the mean velocity and dwell time distribution; nevertheless, we expect that more valuable information is embedded in the trajectory.
Despite the fact that our method requires only the single-molecule trajectory and the assumptions of the balance condition for the probe's motion in each chemical state, the Markovian properties, and the number of states, it provides information about the state-specific potential profile, mechanochemical coupling, and energetics.
The method demonstrated that the mechanochemical picture of the torque generation is elucidated in terms of the basic concepts such as ``power stroke'' and ``diffusion catch'' and that the position-dependent substrate binding is the key feature of the mechanism behind the efficient free-energy conversion of the F$_1$-motor.
The method is simple and will provide a promising methodology for understanding the properties of not only molecular motors but also systems in a broad range of fields.

\section{Materials and methods}

The experimental system was essentially the same as the previous studies\cite{Toyabe2010a, Toyabe2011}.
Rotation of F$_1$-motor molecules ($\alpha_3\beta_3\gamma$ sub complex) derived from a thermophilic {\it Bacillus} PS3 was probed by attaching streptavidin-coated polystyrene particles (diameter = 0.3 $\mu$m, Seradyn) to the biotylated $\gamma$-shaft.
Single-molecule observation was performed in the presence of 0.5 $\mu$M MgATP, 2.5 $\mu$M MgADP, and 1 mM \PP in a buffer containing 5 mM MOPS-K and 1 mM MgCl$_2$ on a microscope (Olympus) at 9,000 Hz (exposure time = 40 $\mu$s) using a 100$\times$ objective (NA = 1.49, Olympus), a Xenon lamp, and a high-speed camera (Basler) at a room temperature (25 $\pm$ 1 $^\circ\!$C).
The trajectory consists of approximately 2,500,000 frames for each molecule typically.
The mean rotational velocity and the diffusion coefficient were $1.9 \pm 0.9$ Hz and $14.3 \pm 3.6$ rad$^2/$s, respectively (mean $\pm$ SD, 36 molecules).
The data including long pause presumably due to the MgADP-inhibited state are excluded from the analysis.
Bin width for calculating $U_\mathrm{i}$ was 1$^\circ$.
The value of the standard free energy change of an ATP hydrolysis ($\Delta\mu^\circ$) depends on literatures.
We took the average of $\Delta\mu^\circ$ calculated on the basis of three references \cite{Panke1997, Guynn1973, Rosing1972} and calculated $\Delta\mu$ \cite{Krab1992, Toyabe2010a, Toyabe2011}:
\begin{equation}
\Delta\mu=\Delta\mu^\circ+k_\mathrm{B}T\ln\mathrm{\frac{[ATP]}{[ADP][P_i]}} .
\end{equation}

\acknowledgments
We appreciate Masaki Sano, Takahiro Sagawa, and Kyogo Kawaguchi for helpful discussions.
This work was supported by Japan Science and Technology Agency (JST) and Grant-in-Aid for Scientific Research.
S.T. was supported by Alexander von Humboldt Foundation.

\bibliographystyle{eplbib}
\bibliography{epl}

\begin{thebibliography}{10}
\expandafter\ifx\csname url\endcsname\relax\def\url#1{\texttt{#1}}\fi

\bibitem{Boyer1993ah}
\Name{Boyer P.~D.} \REVIEW{Biochim. Biophys. Acta }{1140}{1993}{215}.

\bibitem{Abrahams1994}
\Name{Abrahams J.~P., Leslie A. G.~W., Lutter R. \and Walker J.~E.}
  \REVIEW{Nature }{370}{1994}{621}.

\bibitem{Noji1997}
\Name{Noji H., Yasuda R., Yoshida M. \and Kinosita, Jr. K.} \REVIEW{Nature
  }{386}{1997}{299}.

\bibitem{Yasuda1998}
\Name{Yasuda R., Noji H., Kinosita, Jr. K. \and Yoshida M.} \REVIEW{Cell
  }{93}{1998}{1117}.

\bibitem{Toyabe2010a}
\Name{Toyabe S., Okamoto T., Watanabe-Nakayama T., Taketani H., Kudo S. \and
  Muneyuki E.} \REVIEW{Phys.\ Rev.\ Lett. }{104}{2010}{198103}.

\bibitem{Hayashi2010}
\Name{Hayashi K., Ueno H., Iino R. \and Noji H.} \REVIEW{Phys.\ Rev.\ Lett.
  }{104}{2010}{218103}.

\bibitem{Toyabe2011}
\Name{Toyabe S., Watanabe-Nakayama T., Okamoto T., Kudo S. \and Muneyuki E.}
  \REVIEW{Proc. Nat. Acad. Sci. USA }{108}{2011}{17951}.

\bibitem{Adachi2007}
\Name{Adachi K., Oiwa K., Nishizaka T., Furuike S., Noji H., Itoh H., Yoshida
  M. \and Kazuhiko~Kinosita J.} \REVIEW{Cell }{130}{2007}{309}.

\bibitem{Watanabe2010}
\Name{Watanabe R., Iino R. \and Noji H.} \REVIEW{Nature Chem. Biol.
  }{6}{2010}{814}.

\bibitem{Masaike2008}
\Name{Masaike T., Koyama-Horibe F., Oiwa K., Yoshida M. \and Nishizaka T.}
  \REVIEW{Nature Str. Mol. Biol. }{15}{2008}{1326}.

\bibitem{Gaspard2007}
\Name{Gaspard P. \and Gerritsma E.} \REVIEW{J. Thor. Biol. }{247}{2007}{672}.

\bibitem{Risken}
\Name{Risken H.} \Book{The {Fokker-Planck} Equation} (Springer, Berlin) 1984.

\bibitem{NumericalRecipes3}
\Name{Press W.~H., Teukolsky S.~A., Vetterling W.~T. \and Flannery B.~P.}
  \Book{Numerical Recipes Third Edition} (Cambridge University Press,
  Cambridge) 2007.

\bibitem{Adachi2010}
\Name{Adachi K., Furuike S., Hossain M.~D., Kazuhiko~Kinosita J., Onoue Y. \and
  Shimo-Kon R.} \REVIEW{Springer Series in Chem. Phys. }{96}{2010}{271}.

\bibitem{Toyabe2010b}
\Name{Toyabe S., Sagawa T., Ueda M., Muneyuki E. \and Sano M.}
  \REVIEW{Nature\,Phys. }{6}{2010}{988}.

\bibitem{Sekimoto2010}
\Name{Sekimoto K.} \Book{Stochastic Energetics (Lecture Notes in Physics)}
  (Springer, Berlin) 2010.

\bibitem{Harada2005}
\Name{Harada T. \and Sasa S.-I.} \REVIEW{Phys. Rev. Lett. }{95}{2005}{130602}.

\bibitem{Yasuda2001}
\Name{Yasuda R., Noji H., Yoshida M., Kinosita, Jr. K. \and Itoh H.}
  \REVIEW{Nature }{410}{2001}{898}.

\bibitem{Shimabukuro2003}
\Name{Shimabukuro K., Yasuda R., Muneyuki E., Hara K.~Y., Kazuhiko~Kinosita J.
  \and Yoshida M.} \REVIEW{Proc. Nat. Acad. Sci. USA }{100}{2003}{14731}.

\bibitem{Prost1994}
\Name{Prost J., Chauwin J.-F., Peliti L. \and Ajdari A.} \REVIEW{Phys.\ Rev.\
  Lett. }{72}{1994}{2652}.

\bibitem{Wang2003}
\Name{Wang H.} \REVIEW{IEE Proc.-Nanobiotechnol }{150}{2003}{128}.

\bibitem{Ueno2010}
\Name{Ueno H., Nishikawa S., Iino R., Tabata K.~V., Sakakihara S., Yanagida T.
  \and Noji H.} \REVIEW{Biophys. J. }{98}{2010}{2014}.

\bibitem{Panke1997}
\Name{P\"anke O. \and Rumberg B.} \REVIEW{Biochim. Biophys. Acta
  }{1322}{1997}{183}.

\bibitem{Guynn1973}
\Name{Guynn R.~W. \and Veech R.~L.} \REVIEW{J. Biol. Chem. }{248}{1973}{6966}.

\bibitem{Rosing1972}
\Name{Rosing J. \and Slater E.~C.} \REVIEW{Biochim. Biophys. Acta
  }{267}{1972}{275}.

\bibitem{Krab1992}
\Name{Krab K. \and {van Wezel} J.} \REVIEW{Biochim. Biophys. Acta
  }{1098}{1992}{172}.

\end{thebibliography}

\end{document}